\documentclass{article}

\usepackage{arxiv}

\usepackage[utf8]{inputenc} 
\usepackage[T1]{fontenc}    
\usepackage{hyperref}       
\usepackage{url}            
\usepackage{booktabs}       
\usepackage{amsfonts}       
\usepackage{nicefrac}       
\usepackage{microtype}      
\usepackage{lipsum}		
\usepackage{graphicx}
\usepackage{natbib}
\usepackage{doi}

\title{Generalization Across Experimental Parameters in Machine Learning Analysis of High Resolution Transmission Electron Microscopy Datasets}

\author{{Katherine Sytwu} \\
	Molecular Foundry\\
	Lawrence Berkeley National Laboratory\\
	Berkeley, CA 94720 \\
	\texttt{ksytwu@lbl.gov} \\
	\And
	{Luis Rangel DaCosta} \\
	Department of Materials Science and Engineering\\
	University of California Berkeley\\
	Berkeley, CA 94720 \\
	\texttt{} \\
    Molecular Foundry\\
	Lawrence Berkeley National Laboratory\\
	Berkeley, CA 94720 \\
	\And
	{Mary C. Scott} \\
	Department of Materials Science and Engineering\\
	University of California Berkeley\\
	Berkeley, CA 94720 \\
	\texttt{} \\
	Molecular Foundry\\
	Lawrence Berkeley National Laboratory\\
	Berkeley, CA 94720 \\
	\texttt{MCScott@lbl.gov} \\
}

\date{}

\renewcommand{\shorttitle}{HRTEM Generalization}

\begin{document}

\maketitle

\begin{abstract}

Neural networks are promising tools for high-throughput and accurate transmission electron microscopy (TEM) analysis of nanomaterials, but are known to generalize poorly on data that is ``out-of-distribution" from their training data. Given the limited set of image features typically seen in high-resolution TEM imaging, it is unclear which images are considered out-of-distribution from others. Here, we investigate how the choice of metadata features in the training dataset influences neural network performance, focusing on the example task of nanoparticle segmentation. We train and validate neural networks across curated, experimentally-collected high-resolution TEM image datasets of nanoparticles under controlled imaging and material parameters, including magnification, dosage, nanoparticle diameter, and nanoparticle material. Overall, we find that our neural networks are not robust across microscope parameters, but do generalize across certain sample parameters.  Additionally, data preprocessing heavily influences the generalizability of neural networks trained on nominally similar datasets. Our results highlight the need to understand how dataset features affect deployment of data-driven algorithms. 

\end{abstract}

\section{Introduction}
With increasing amounts of data from faster detector speeds and new automated microscope setups, there is a pressing need for high-throughput analysis of high-resolution transmission electron microscope (HRTEM) images of nanomaterials. HRTEM enables atomic-scale visualization of material structure with high temporal resolution, making it a useful imaging modality for high-throughput and in situ TEM experiments of nanoparticle synthesis and behavior. The most promising HRTEM image analysis methods to date have been based on convolutional neural networks (CNNs), a class of machine learning models that naturally take advantage of spatial correlations in image data \citep{madsen2018deep,vincent2021developing,groschner2021machine}. These algorithms utilize a framework in which patterns and trends are extracted from a large corpus of data, called the training set, and then evaluated on data the algorithm has not seen during training. The subsequent performance then depends on both the construction of a suitable optimization problem (which depends on the network architecture, training data, and loss function) and the procedure used to solve for the optimal parameters.

While CNNs have consistently outperformed traditional image analysis methods, CNNs and other machine learning models have also been empirically shown to not perform as well on data that is separate from their training set \citep{pmlr-v97-recht19a,torralba2011unbiased}. This inability to generalize has consequences for deploying CNNs for large-scale microscopy analysis, for instance in determining which networks are reusable across multiple experiments or reliable for data streams with changing conditions, like in situ data. Generalization issues are typically categorized in two ways: 1) in-distribution generalization, or the ability to generalize on data that has been nominally sampled from a similar distribution as the training data and whose drop in performance is commonly referred to as the ``generalization gap", and 2) out-of-distribution generalization, or the ability to extrapolate to new data that is known to be different from the training set. While there has been a growing amount of research focused on algorithmic solutions to minimize generalization issues \citep{shen2021towards}, we first need to understand under what conditions generalization problems occur. Such an analysis requires domain-specific knowledge which associates model performance gaps with domain-knowledge of the modified image or data features \citep{kaufmann2021acquisition, liu2020neural,li2023critical}.

With HRTEM data, it is unclear what types of images are considered out-of-distribution from others. While metadata information like sample and/or imaging parameters may designate images as different from one another, it is unknown whether a trained neural network would be sensitive to such changes given the limited number of image features typically seen in HRTEM images. There is also limited knowledge on how the training dataset affects neural network performance, despite our (often) relatively complete understanding of both the sample and imaging process. While there have been some attempts to understand the effect of the training dataset with simulated data \citep{vincent2021developing}, we lack experimental benchmarks to fully validate these generalization effects. With more data-driven models being proposed and developed by the microscopy community, there is a need to understand the reusability of these models on new datasets, and under what conditions they succeed or fail \citep{wei2023benchmark, larsen2023quantifying}.

\begin{figure}
    \centering
    \includegraphics{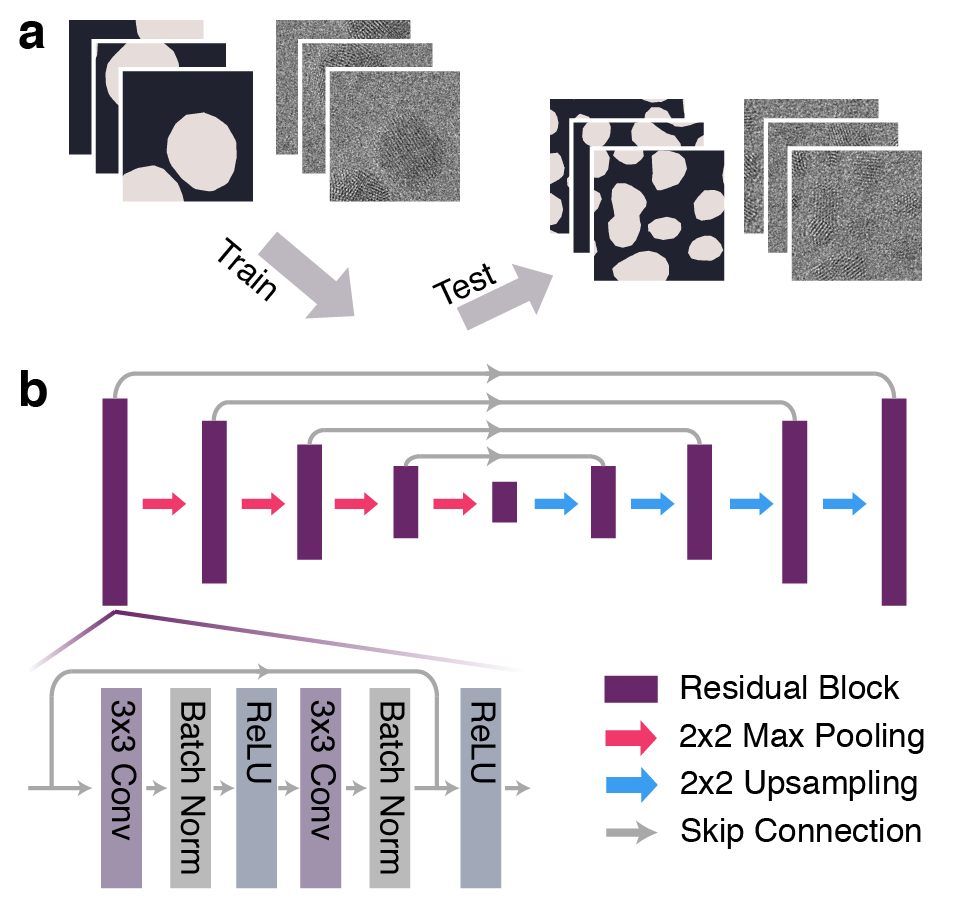}
    \caption{Overview of the network training and testing protocols. (a) Datasets with specified metadata parameters are labeled and created from large HRTEM images and used to train and test neural networks. (b) The residual UNet neural network architecture used for all models in this paper. }
    \label{fig:architecture}
\end{figure}

In this paper, we systematically examine the robustness of neural networks trained to identify nanoparticles in HRTEM images (Figure \ref{fig:architecture}a), focusing on the effect of microscope and sample parameters in the training set, including magnification, electron dosage, nanoparticle diameter, and nanoparticle material. As an example task, we focus on segmentation, or pixel-wise classification, of atomically-resolved crystalline nanoparticles against an amorphous background, a typical initial image processing step for further analysis of atomic defects or crystal structure \citep{groschner2021machine}, or nanoparticle dynamic behavior \citep{yao2020machine}. By curating experimental HRTEM datasets with controlled imaging and sample parameters, we not only qualitatively identify conditions under which we expect networks to generalize (or not), but also provide new datasets with extensive metadata that enable benchmarking HRTEM image analysis methods under specified microscopy conditions. In addition to our observations on training set effects, we demonstrate how data preprocessing influences generalization, providing a case study in preparing and utilizing experimental TEM data.

\section{Methods}
\subsection{Sample Preparation}
2.2nm Au nanoparticles with citrate ligands were purchased from Nanopartz. 5nm, 10nm, and 20nm Au nanoparticles capped with tannic acid were purchased from Ted Pella. 5nm Ag nanoparticles with citrate ligands were purchased from nanoComposix. 5nm CdSe nanoparticles with oleylamine ligands were purchased from Strem Chemicals. To create a TEM sample from aqueous nanoparticle solutions (Au, Ag), an ultrathin carbon grid (Ted Pella) was plasma cleaned with a shield for 3 seconds to promote hydrophilicity, then 5 $\mu$L of the purchased nanoparticle solution was dropcast onto the grid, let sit for 5 minutes, and excess liquid was wiped off with a Kimwipe. For the CdSe nanoparticles, the nanoparticle solution was diluted to 0.625\% of the original concentration with hexane, and 5$\mu$L of the diluted nanoparticle solution was dropcast onto an ultrathin carbon grid (Ted Pella) and let evaporate. 
\subsection{TEM Imaging}
HRTEM images were taken with a TEAM 0.5 aberration-corrected microscope operated at 300kV and a OneView camera (Gatan) at full resolution (4096 x 4096 pixels). 
\subsection{Preprocessing and Dataset Creation}
All HRTEM images were labeled by hand into segmented images using Labelbox. To create a dataset, raw images (and their corresponding labels) were selected from the larger data repository using metadata (i.e. microscope conditions, nanoparticle parameters, etc.) , and then preprocessed into a dataset \citep{sytwu2023segmentation}. Preprocessing consisted of four steps: 1) Removal of x-rays. 2) Flat-field correction. 3) Image value rescaling. 4) Divide into smaller patches. We apply all preprocessing steps by image to ensure that our methods scale with new data (i.e. adding more images to a dataset) and are reflective of model deployment, which is likely to be done by image. X-rays were removed by averaging the surrounding pixels of outlier points above a certain threshold (1500 counts) above the mean counts. For flat field correction, we estimate the uneven illumination using iterative weighted linear regression to a 2D Bezier basis (n=2, m=2) \citep{sadre2021deep}, and divide out the estimated illumination profile. The iterative reweighting lessens the contribution from nanoparticle regions such that the substrate regions are primarily used to determine the uneven illumination. The pixel values of each image are then rescaled using either normalization (set minimum to 0 and maximum to 1), standardization (set mean to 0 and standard deviation to 1), or a histogram-based scaling procedure similar to Digital Micrograph (normalize, but ignore the pixels outside the 1st and 99th percentiles). Finally, images are divided into 512x512 pixel patches to reduce GPU memory requirements during network training and patches that are mostly substrate are removed to obtain better class balance. The datasets used in this paper are described in more detail in Table S1. 
\subsection{Neural Network Training and Testing}
For every dataset, five networks are trained using five-fold cross-validation to account for the variations from initial conditions and choice of test set. Networks are trained using a 70-10-20 percentage train-validation-test set split. After removing the test set, 1/8th of the remaining images are held as a validation set. Patches are assigned sequentially such that it is less likely for patches from similar image regions to end up in both the training and test sets. Each training/validation/test dataset is then augmented using the eight dihedral augmentations and shuffled.

Our neural network model architecture is a residual variant of the UNet architecture \citep{ronneberger2015u} with four residual blocks and detailed more in \cite{sytwu2022understanding} (Figure \ref{fig:architecture}b). Models are trained under a supervised learning framework, using cross-entropy loss with a learning rate of 1e-4 with an Adam optimizer (default parameters). During training, we additionally augment 50\% of the images with random rotations between 0 and 360 degrees which empirically produces smoother prediction edges, but do not apply these random rotations to the validation or test set. We train for 250 epochs and save the model weights with the lowest validation loss within those 250 epochs. All training is done locally on a NVIDIA RTX3090 GPU. 

We evaluate our models using the hard dice score, also known as the F1 score, which quantifies the similarity between the prediction and expert-provided label. The hard dice score can be calculated by $\frac{2TP}{2TP + FP+FN}$ for a binary classification, and ranges between 0 (for complete disagreement) to 1 (for exact agreement). The results reported in this paper are the mean and standard deviation of the 5 trained models on either the test set (if drawn from the same dataset) or the entire other datasets.  

\section{Results}
\subsection{Preprocessing}

\begin{figure*}
    \centering
    \includegraphics{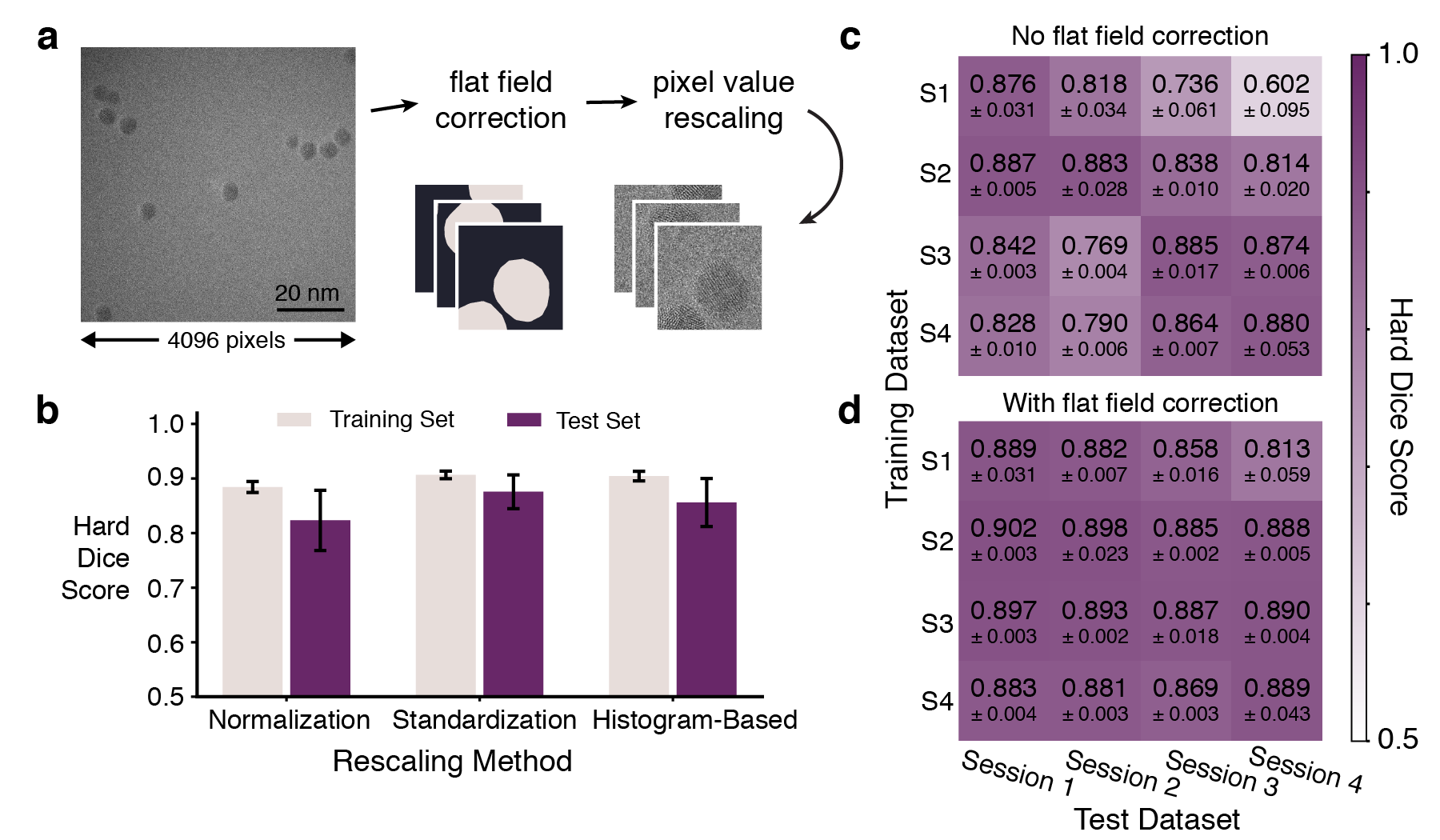}
    \caption{Effect of data preprocessing on network generalizability. (a) Overview of the data preprocessing workflow, from camera output to dataset creation. (b) The effect of pixel value rescaling procedures (normalization, standardization, and histogram-based) on the average performance of the training set and the test set for networks trained on the same 5nm Au nanoparticle dataset. Error bars refer to standard deviation over 5 networks. (c,d) Confusion matrices of network performance when trained and tested on images of 5nm Au nanoparticles taken at 0.02nm/pixel scale and 423 e/A\textsuperscript{2} dosage from four different sessions (c) without any flat field correction and (d) with flat field correction. Error refers to standard deviation over 5 networks.}
    \label{fig:preprocessing}
\end{figure*}

In order to identify the effect of training dataset on network performance, models need to first generalize well on images with nominally similar microscope conditions and sample parameters, whether the images are taken from the same dataset or during a different microscope session. We find that choices in data preprocessing highly affect whether this statement holds true. Preprocessing encompasses the conversion process from raw camera data in the form of CCD counts into a data format that is conducive for neural network training. There are two types of preprocessing steps: one is related to how datasets are created from the acquired data such that we can train neural networks in a memory-efficient and class-balanced manner (i.e. dividing large images into smaller patches that we feed into the neural network during training); the other is related to how image data values are converted into a standard format that enables generalization across quantitatively different camera outputs (i.e. if recorded counts are slightly different from changes in the camera gain). In this section, we will focus on the latter type of data preprocessing, and its implications for network generalization, with the key steps highlighted in Figure \ref{fig:preprocessing}a. 

Pixel value rescaling is necessary to convert output camera data into a standard format that is robust against exact electron counts. TEM image data is outputted as an array of counts, often from a high dynamic range sensor, whose exact pixel values correspond to detector and microscope parameters like gain and exact electron dosage. We test three different rescaling methods: normalization, standardization, and a histogram-based scaling method similar to Digital Micrograph, a commonly used micrograph viewing software. Given a single set of images taken during the same session (8 images of 5nm Au nanoparticles, which results in 211 patches), we create three datasets that have the same content, but only differ by how the data is rescaled.  

We find that the choice of pixel value rescaling method affects the generalization gap, or how well networks generalize to new images from the same microscope session. The rescaling method does not seem to noticeably affect the network's ability to converge to a solution, as evidenced by the high dice scores and low standard deviation of the training set performance for all 3 rescaling methods, but does affect generalization performance to the test set (Figure \ref{fig:preprocessing}b). Normalization is the least robust rescaling method, having both the largest drop and variation in average test set performance relative to training set performance. Both of these trends suggest that by normalizing images, network performance is influenced by the sampling of the test set.  

We attribute the performance differences across pixel rescaling methods to the larger variations in image values when normalizing, compared to more consistent nanoparticle contrast and background values when standardizing or undergoing histogram-based rescaling. TEM images often do not use the full dynamic range of the scientific sensor, and so normalization is sensitive to fluctuations in the long tail of pixel value counts. For small datasets, these variations in image contrast across images can lead to large differences between the training and validation/test sets, while standardization and histogram-based rescaling result in more consistent pixel value distributions between images (Figure S1). Due to standardization's highest performance and lowest variance, we standardize HRTEM image data for the subsequent datasets used in this paper. We do note that standardization partially relies on the assumption that the image values are normally distributed. This assumption holds mostly true for a wide-view image where the majority of the image area is amorphous substrate, but can potentially fail for an image whose majority area is crystalline material with a bimodal pixel value distribution from strongly diffracting lattice fringes. 

In addition to consistent performance across a single microscope session or dataset, a robust algorithm should also be consistent across datasets that are nominally similar. We test our neural networks’ ability to generalize to datasets taken during four different microscope sessions but with nominally similar sample and imaging conditions (5nm Au nanoparticles taken at 0.02nm pixel scale with 423 e/\AA\textsuperscript{2} dosage). Figure \ref{fig:preprocessing}c shows a confusion matrix of the networks’ performance, with the diagonal elements highlighting the performance on test data taken from the same dataset, and the off-diagonal elements showcasing the performance on data from sessions different from the training dataset. These networks primarily perform well on test sets drawn from the same dataset (i.e. same microscope session) they were trained on, but fail to generalize to nominally similar data, suggesting that there is session-dependent information that the models are capturing. 

By applying flat field correction to our images, we are able to obtain better generalization performance across microscope sessions. This preprocessing step corrects for uneven illumination across the image caused by either shifts in the monochromator or incorrect gain references. As seen in Figure \ref{fig:preprocessing}d, once the images are flat field corrected, networks generalize much better to other sessions of nominally similar data. Flat field correction is particularly influential in our datasets because preprocessing is done per-image; the correction ensures that there is less variation across patches such that patch statistics better match larger scale image statistics. Note that flat field correction does not seem to impact how well the networks analyze the data–-the diagonal elements of the confusion matrix retain similar performance regardless of flat field correction. Therefore, flat field correction primarily removes session-dependent experimental artifacts that affect generalization.

\subsection{Generalizability across microscope parameters} 
\begin{figure*}[!h]
    \centering
    \includegraphics[width=\textwidth]{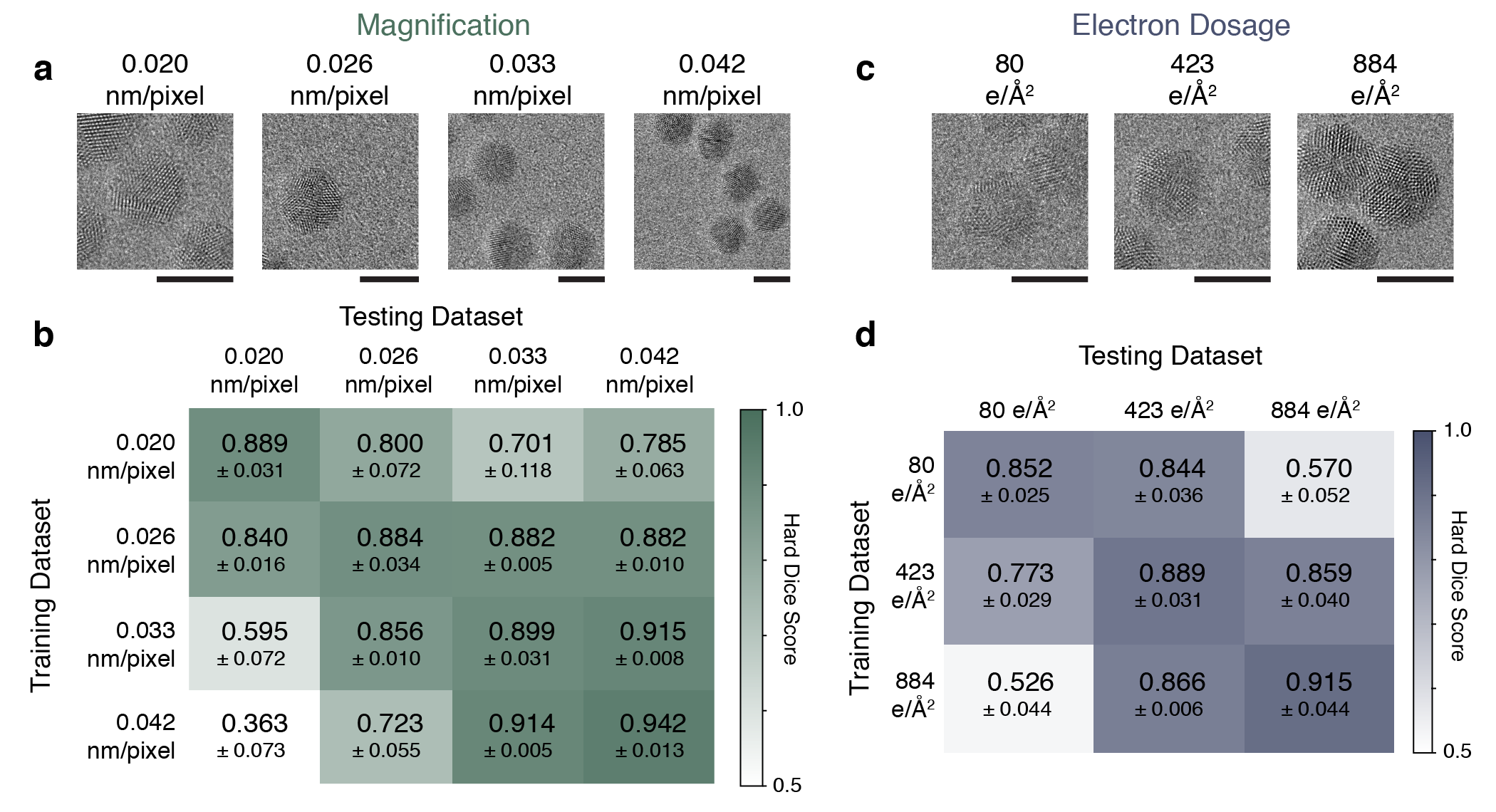}
    \caption{Network generalizability over microscope conditions. (a) Sample images from the four datasets of 5nm Au nanoparticles taken at different microscope magnifications. (b) Confusion matrix of network performance when trained and tested with datasets taken at different magnifications. (c) Sample images from the three datasets of 5nm Au nanoparticles taken at various dosage conditions.  (d) Confusion matrix of network performance when trained and tested with datasets taken at different electron dosages. All scale bars are 5nm.}
    \label{fig:microscope_generalization}
\end{figure*}

Microscope parameters heavily affect how a HRTEM image is formed and the subsequent observed image features. As a sanity check, we first investigate the generalizability of networks across microscope magnifications. Our networks are not expected to generalize well since lattice fringes, a key nanoparticle image feature, have a characteristic length scale and our CNNs are, by construction, not scale-invariant. We create four datasets, each of 5nm Au nanoparticles taken at similar dosages but different magnifications (Figure \ref{fig:microscope_generalization}a), and then train and test networks across the four datasets. As expected, neural network performance is worse on images taken at a different magnification than the training dataset, with a larger drop in performance on test sets with a greater difference in pixel scale (Figure \ref{fig:microscope_generalization}b).

From the confusion matrix, we see that generalization behavior is not necessarily symmetrical. For instance, networks trained on images taken at 0.042nm pixel scale perform extremely poorly on images taken at 0.02nm pixel scale, but this difference in performance is smaller vice versa. We hypothesize that this asymmetry is from the neural network using additional information beyond just the spatial frequency of lattice fringes to make its decisions \citep{sytwu2022understanding}. In HRTEM imaging, changing the magnification alters the relative contributions of amplitude and diffraction (phase) effects in image formation, alongside rescaling the image. Nanoparticles have greater image contrast in images taken at lower magnifications (i.e. 0.042 nm/pixel) than in images taken at higher magnification (0.02 nm/pixel), and are thus  qualitatively easier to detect in low magnification images than higher ones. The ease of distinguishing between nanoparticle and background in the lower magnification images is also noted by the overall higher performance on the 0.042 nm/pixel dataset.

In addition to magnification, we find that networks do not generalize well to datasets taken at different electron dosages, which affects the signal-to-noise ratio in the image. We again create three datasets, each of 5nm Au nanoparticles taken at 0.02nm pixel scale, but at three different dosages to represent a low dose dataset (80 e/\AA\textsuperscript{2}), a medium dose dataset (423 e/\AA\textsuperscript{2}), and a high dose dataset (884 e/\AA\textsuperscript{2}) (Figure \ref{fig:microscope_generalization}c). Here, we see a slightly more symmetrical confusion matrix, with all networks dropping in performance when tested on data taken at a dosage different from the training dataset (Figure \ref{fig:microscope_generalization}d). Upon further analysis, we observe that networks tested on an image taken at a higher dosage (relative to the training set) tend to oversegment, as evidenced by a higher false positive rate, while networks tested on images taken at a lower dosage tend to undersegment (Figure S4). This suggests that evaluating on a dataset with a dosage different from the training dataset could incorrectly bias subsequent nanoparticle size analysis, but more detailed studies with a wider range of dosage values are needed to quantify these potential errors. 

\subsection{Generalization across sample parameters}

\begin{figure*}
    \centering
    \includegraphics{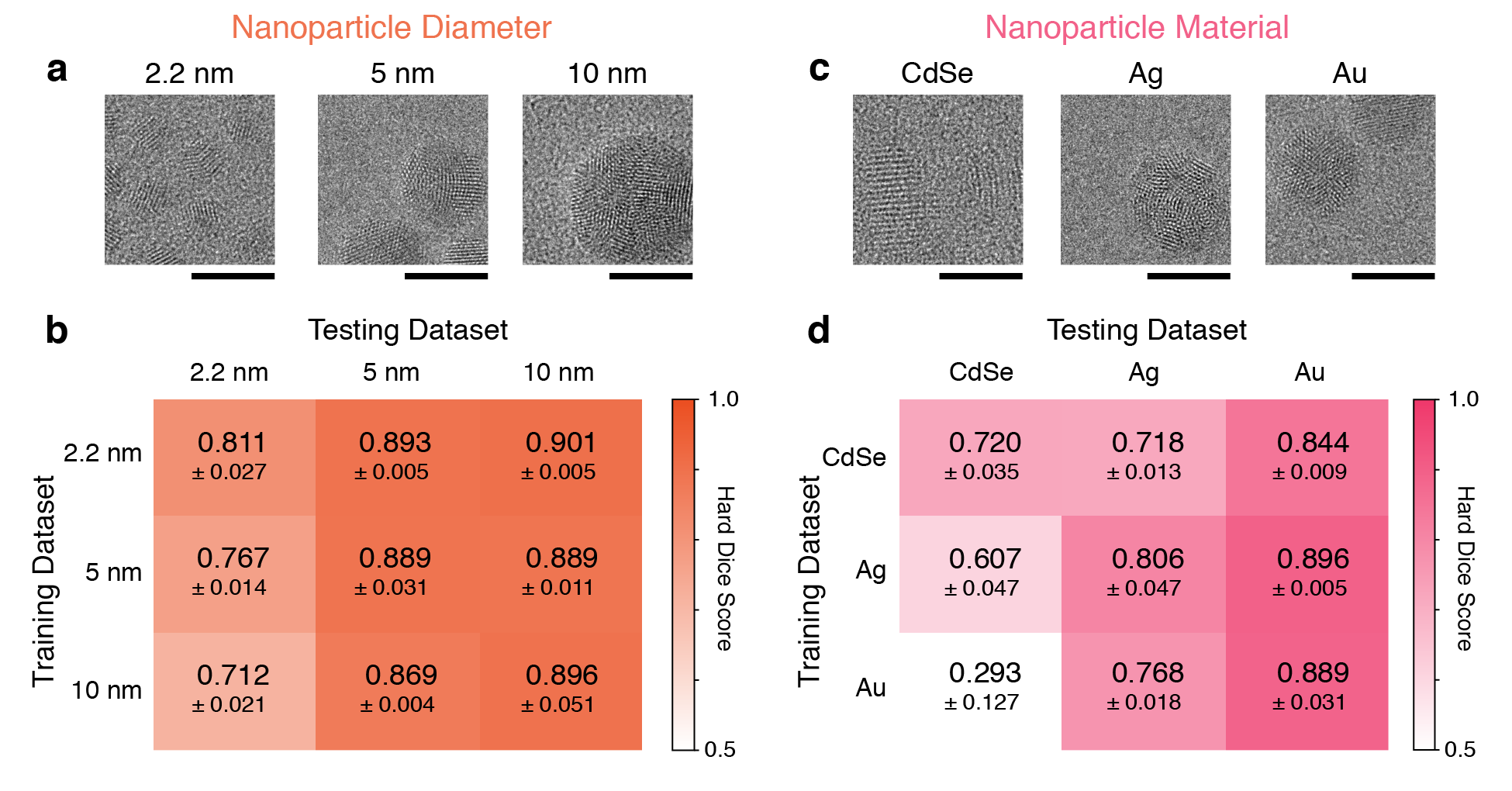}
    \caption{Network generalizability over nanoparticle sample parameters. (a) Sample images from the three datasets of Au nanoparticles of various diameters taken with similar microscope conditions. (b) Confusion matrix of network performance when trained and tested with datasets of Au nanoparticles with different diameters. (c) Sample images from the three datasets of approximately 5nm nanoparticles of either CdSe, Ag, or Au taken with similar microscope conditions. (d) Confusion matrix of network performance when trained and tested with datasets of nanoparticles of different materials. All scale bars are 5nm.}
    \label{fig:sample_generalization}
\end{figure*}

Understanding the reliability of a trained neural network across sample parameters is especially crucial for in situ studies and automated microscopy where microscope parameters are usually fixed but sample parameters may change or be unknown in the future. For nanoparticle datasets, one commonly varying sample parameter is nanoparticle size. We again create three datasets, each of Au nanoparticles taken at similar microscope conditions but varying in average nanoparticle diameter from 2.2nm to 10nm (Figure \ref{fig:sample_generalization}a). While the observed lattice fringes have the same characteristic length scale in all three datasets, larger nanoparticles are thicker and therefore have greater nanoparticle contrast (both amplitude and phase-contrast) against the substrate background. When evaluating network generalization (Figure \ref{fig:sample_generalization}b), we find that some models and datasets generalize well. All models perform equally well on the 5nm and 10nm datasets, but there is some variation in performance on the 2.2nm dataset depending on training dataset. All models perform worse on the 2.2nm dataset, likely due to the low nanoparticle contrast and difficulty of interpreting the images. Qualitative analysis of the predicted labels of the 2.2nm dataset suggest that the lower dice score may also be from the network identifying particles that were missed by the human labeler (Figure S5). When normalizing for dataset difficulty, it is clearer that models trained on 2.2nm data generalize better than models trained on larger nanoparticles (Figure S2). These results suggest that networks could be trained to perform well on image data streams without needing to know the exact nanoparticle size beforehand.

In addition to nanoparticle size, nanoparticles can also vary in their material, which leads to differences in contrast (from atomic number, Z), and nanoparticle lattice features (from lattice spacing and crystal structure). We create three datasets of approximately 5nm nanoparticles taken at similar microscope conditions, but varying in material: Au, Ag and CdSe. Au and Ag are both fcc metals with similar lattice spacings, but differ in contrast (Z\textsubscript{Au} = 79, Z\textsubscript{Ag} = 47). CdSe nanoparticles, on the other hand, can take on either a wurtzite (hexagonal) or zinc blende (fcc) structure (both appear in our sample) with average lattice spacings greater than Au and Ag, but with contrast similar to Ag (Z\textsubscript{Cd} = 48). Again, we see an imbalance in network performance depending on the dataset, with both Au- and Ag-trained networks performing well on the Au datasest, and a strong dependence on training data for the CdSe and Ag datasets (Figure \ref{fig:microscope_generalization}d). For the CdSe and Ag datasets, training on similar data does not even provide very high performance. Most interestingly, the CdSe-trained model performs decently well on the Au dataset, despite the CdSe nanoparticle regions having both different contrast and frequency information from the Au nanoparticle regions. 

\section*{Discussion}
Overall, we find that there is potential for networks to generalize under certain sample parameters (nanoparticle size and material) but not over different microscope parameters (magnification and dosage). This suggests that pre-trained neural networks could be used for data streams with controlled imaging parameters, for instance with in situ datasets and automated microscopy. We also find that networks trained on more difficult-to-interpret data tend to generalize to new data better than networks trained on easier-to-interpret data, which can be observed in most of our confusion matrices. The datasets have been qualitatively ordered from lowest to highest in terms of how easily the nanoparticles are distinguishable, with higher nanoparticle contrast and observable lattice fringes making an image easier to interpret. Consistently, the generalization performance is worse in the lower left corner of our confusion matrices (train on easy images, test on harder images) compared to the upper right corner (train on hard images, test on easier images). Since labeling difficult-to-interpret data is prone to larger human bias and error, these results highlight the need for simulation-based or multimodal datasets with accurate ground truth information to create useful training data \citep{madsen2018deep,vincent2021developing}. 

\begin{figure*}
    \centering
    \includegraphics{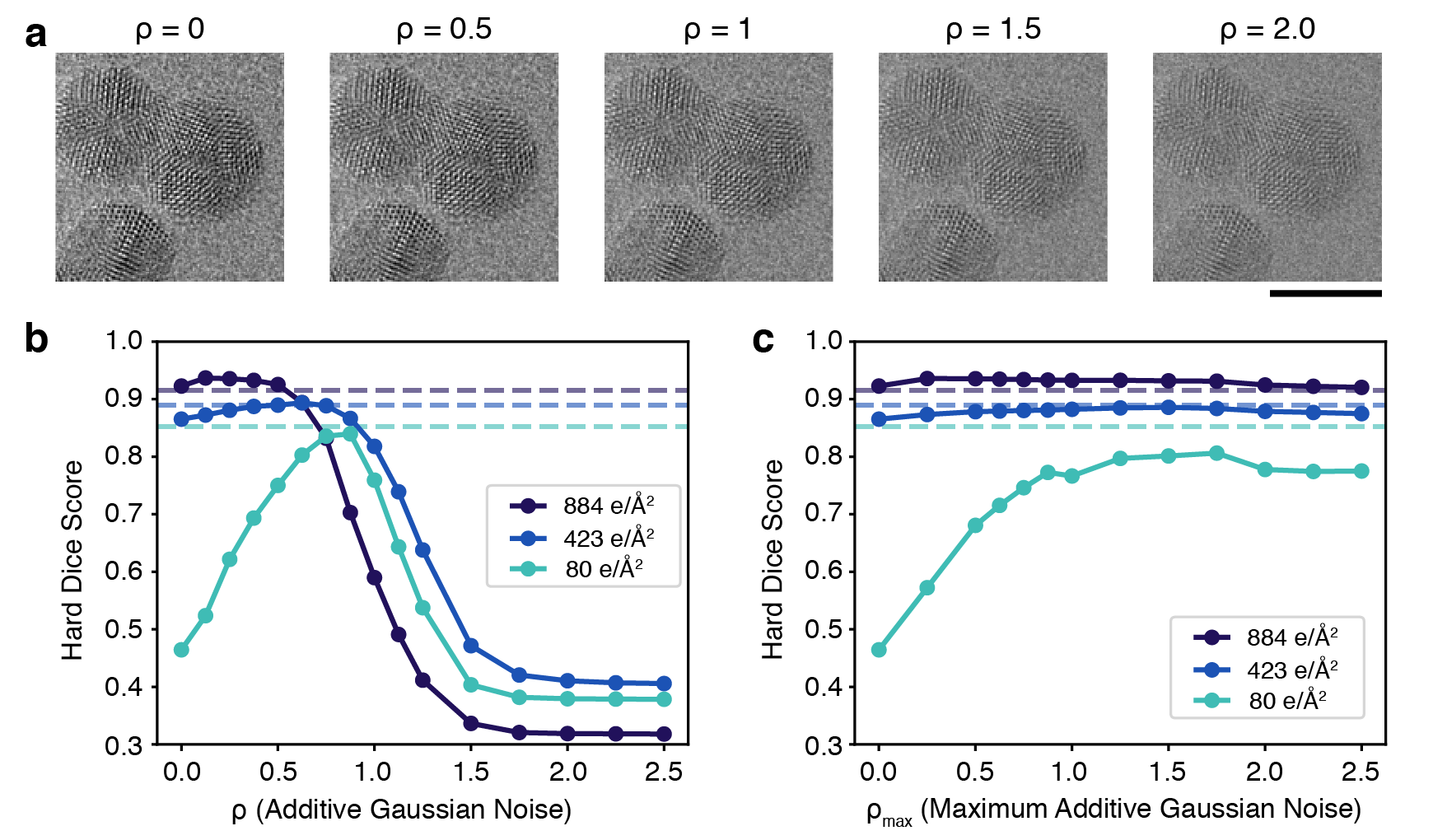}
    \caption{The effect of additive Gaussian noise on experimental data. (a) Sample image from the 884 e/\AA\textsuperscript{2} dataset (same as in Figure \ref{fig:microscope_generalization}c with various amounts of additive Gaussian noise of scale $\rho$. Scale bar is 5nm. (b,c) Performance of neural networks trained on the 884 e/\AA\textsuperscript{2} dataset augmented with (b) additive Gaussian noise of scale $\rho$ or (c) additive Gaussian noise with scale sampled between $[0,\rho_\textsubscript{max}]$ when tested on the 884 e/\AA\textsuperscript{2} test set (dark blue), 423 e/\AA\textsuperscript{2} dataset (blue), and 80 e/\AA\textsuperscript{2} dataset (turquoise). Dotted lines indicate the average performance of the respective dataset when trained on images from the same dataset. }
    \label{fig:noise_augmentation}
\end{figure*}

In the absence of collecting more data to improve the generalizability of our networks, we can alternatively mimic lower contrast and more difficult to interpret datasets by adding noise and corrupting information in the higher contrast datasets for which we have higher confidence in the labeling. Upon adding Gaussian noise to the images during training, we lower the nanoparticle contrast, but retain the lattice fringe features that denote nanoparticle regions (Figure \ref{fig:noise_augmentation}a). Note that additive Gaussian noise augmentation is a known regularization protocol to prevent overfitting \citep{bishop1995training} and synthetically promote robustness \citep{gilmer2019adversarial}.

As an example, we explore how additive noise augmentation affects generalizability across electron dosage. We train a series of models such that their training dataset of high dosage images (884 e/\AA\textsuperscript{2}) is augmented with additive Gaussian noise with a standard deviation of $\rho$. We then evaluate the performance of these noise-augmented models on the original 884 e/\AA\textsuperscript{2} test set (high dose), 423 e/\AA\textsuperscript{2} dataset (medium dose), and 80 e/\AA\textsuperscript{2} dataset (low dose). As seen in Figure \ref{fig:noise_augmentation}b, performance on all three datasets improve upon additive Gaussian noise augmentation, though the ideal amount of additive noise $\rho$ depends on dataset. As expected, more additive noise is needed to improve performance on lower dosage datasets. Additionally, for all datasets, additive Gaussian noise augmentation helps networks meet or exceed the average performance of neural networks trained on experimentally-collected similar data. This is surprising given that the measured noise from the OneView camera follows a scaled Poissonian distribution and not a Gaussian. It is unclear whether the high performance from this augmentation is from matching dataset characteristics or from regularizing decision boundaries. The optimal augmented noise level does not match the experimentally-collected dataset in either nanoparticle contrast (by matching histogram medians), nor noise statistics (by matching image roughness) (Figure S7). However, when repeating this noise augmentation procedure on the medium-dose dataset, the noise-augmented models generalize poorly to higher dose data and require less additive noise to generalize well to lower dose data, suggesting that there is some dependence on dataset characteristics (Figure S8). All networks degrade in performance when $\rho > 1$ standard deviation, likely because this large noise augmentation destroys information in the image itself.  

As the necessary additive Gaussian noise scale may not be known a priori, we alternatively set the noise augmentation such that $\rho$ is uniformly sampled between $[0,\rho_\textsubscript{max}]$ during training. Under this augmentation protocol, all noise-augmented-trained models perform well on the high dose and medium dose datasets, but none of them perform well enough on the low dose dataset to compare with low-dose trained models (Figure \ref{fig:noise_augmentation}c). These results suggest that synthetic noise augmentation could be a viable strategy for developing robust networks on HRTEM images with decent signal-to-noise, but does not work effectively to generalize to low dosage images with low signal-to-noise. Recent work has highlighted the need for more accurate noise modeling, especially for low dosage images \citep{larsen2023quantifying}, and our results similarly highlight the difficulty of generalizing to low-dosage images.  

We emphasize that the focus of our results is in the data-driven generalization trends rather than absolute neural network performance, which can be affected by label error and choice of model hyperparameters. As the models in this paper are all trained from hand-labeled experimentally-collected data, there is inherent human bias and error in the labels, primarily at the edges of nanoparticles, which affects the absolute value of the dice scores. Similarly, while our training curves suggest that our networks have converged to a local minima that enables decent performance, there is still room for improvement by fine-tuning both model and optimization hyperparameters. We argue, however, that the generalization trends that we observe are data-dependent and seem to be robust even after hyperparameter tuning; in Figure S6, we show the generalization performance over nanoparticle size after hyperparameter tuning model parameters, and while the overall dice scores are slightly different, the generalization trends are the same as Figure \ref{fig:sample_generalization}b. 

Finally, the observed sensitivity to data preprocessing suggests that we need a closer examination as to how we convert raw scientific data into datasets for machine learning and other data-driven methods. While compressed digital images are easier to share, there needs to be greater transparency on how color mapping was performed, which affects image contrast values, visibility of outliers, and potentially leads to dataset biases \citep{zhong2021study}. Given the generalizability differences due to pixel rescaling method that we see in Figure \ref{fig:preprocessing}b, we recommend that researchers are open about the TEM image creation process, namely how camera data is converted to image data. To this end, we have not only made our processed datasets for all of our models publicly available, but also the raw camera data such that preprocessing steps can be explored \citep{sytwu2023segmentation}. By sharing the raw camera data rather than digital images, we hope to invigorate research into the necessary data preprocessing steps for robust algorithms that work on data from any experiment. 

\section{Conclusions}
We investigated how training dataset creation affects neural network segmentation performance on HRTEM images of nanoparticles. We find that choices in data preprocessing, or the conversion from raw camera data to a machine-learning-ready dataset, heavily impacts the ability for networks to generalize to new data. Overall, we find that our trained neural networks are not generalizable across microscope parameters like magnification and electron dosage, which correspond with changing image features like feature size and signal-to-noise ratio. However, networks are more generalizable across sample parameters like nanoparticle diameter and certain nanoparticle materials, which corresponds with image features like nanoparticle contrast and lattice fringe frequency. These results give insight into the experimental conditions under which we can expect trained neural networks to be reliable, and suggest the varieties of data needed for generalizable neural networks. 

\section{Data Availability}
All processed datasets, raw image data, and corresponding labels used in this paper are available in the Dryad Digital Repository, at https://doi.org/10.7941/D1SP93 \citep{sytwu2023segmentation}. The raw image data is also available at https://portal.nersc.gov/project/m3795/hrtem-generalization/; code to download specific files based on metadata attributes is available on our Github. Code and Jupyter notebooks on dataset creation and model training/testing, trained model weights, and more visualizations of our results are available at https://github.com/ScottLabUCB/HRTEM-Generalization. 

\section{Competing Interests}
The authors declare no competing interests.

\section{Acknowledgements}
K.S. was supported by an appointment to the Intelligence Community Postdoctoral Research Fellowship Program at Lawrence Berkeley National Laboratory administered by Oak Ridge Institute for Science and Education (ORISE) through an interagency agreement between the U.S. Department of Energy and the Office of the Director of National Intelligence (ODNI). L.R.D. was supported by the U.S. Department of Energy, Office of Science, Office of Advanced Scientific Computing Research, Department of Energy Computational Science Graduate Fellowship under Award Number DE-SC0021110. This work was also funded by the US Department of Energy in the program “4D Camera Distillery: From Massive Electron Microscopy Scattering Data to Useful Information with AI/ML". Imaging was done at the Molecular Foundry, which is supported by the Office of Science, Office of Basic Energy Sciences, of the U.S. Department of Energy under Contract No. DE-AC02-05CH11231. 

This report was prepared as an account of work sponsored by an agency of the United States Government. Neither the United States Government nor any agency thereof, nor any of their employees, makes any warranty, express or implied, or assumes any legal liability or responsibility for the accuracy, completeness, or usefulness of any information, apparatus, product, or process disclosed, or represents that its use would not infringe privately owned rights. Reference herein to any specific commercial product, process, or service by trade name, trademark, manufacturer, or otherwise does not necessarily constitute or imply its endorsement, recommendation, or favoring by the United States Government or any agency thereof. The views and opinions of authors expressed herein do not necessarily state or reflect those of the United States Government or any agency thereof.

\bibliographystyle{MandM}

\bibliography{refs}    

\end{document}